\documentclass{iaus260}
\usepackage{graphicx}
\pubyear{2009}
\volume{260}  %% insert here IAU Symposium No.
\pagerange{1--4}
\jname{The R\^ole of Astronomy in Society and Culture}
\editors{D. Valls-Gabaud \& A. Boksenberg, eds.}
\title[Preservice elementary teachers and astronomy]                   %% Give here short title %%
{Some remarks on a current study involving preservice elementary teachers and some basic astronomical phenomena} 
\author[Alejandro Gangui, Mar{\'\i}a Iglesias \& Cynthia Quinteros]           %% Give here short author list %%
{Alejandro Gangui$^{1,2}$, Mar{\'\i}a Iglesias$^2$ \and Cynthia Quinteros$^2$}        %% Give here full author list %%

\affiliation{$^1$Instituto de Astronom{\'\i}a y F{\'\i}sica del Espacio / CONICET, \\ 
                 Ciudad Universitaria, 1428 Buenos Aires, Argentina. \\ 
                 email: {\tt gangui@df.uba.ar} \\[\affilskip]
             $^2$Centro de Formaci\'on e Investigaci\'on en la Ense\~nanza de las Ciencias, \\ FCEyN, Universidad de Buenos Aires.}

\begin{document}
\maketitle

\begin{abstract}
Recent studies have shown that not only primary school students but also their future teachers reach science courses with pre-constructed
and consistent models of the world surrounding them. These ideas include many misconceptions which turn out to be robust and hence make
difficult an appropriate teaching-learning process. We have designed some tools (and show here results with a questionnaire) that proved
helpful in putting in evidence some of the most frequently used alternative models on a few basic astronomical notions. We have tested
this questionnaire with preservice elementary teachers from various normal schools in Buenos Aires and made a first analysis of the
results. The collection of data recovered so far shows that some non-scientific conceptions are indeed part of the prospective teachers'
(scientific) background and, therefore, that the issue deserves special attention during their formal training.  

 %% Add here a maximum of 10 keywords
\end{abstract}

\firstsection % If your document starts with a section,
\section{Context}

In a previous study (Gangui et al., 2008) we showed the results of two pilot tests made in different elementary teacher's schools of the
city of Buenos Aires. We used a questionnaire consisting of twelve questions that ask about eight different topics in basic
astronomy. After having tried this instrument several times, the final version of it was used with two groups of preservice elementary
teachers. The total number (n) of inquired people with the new questionnaire was 51 students. However, some of the questions did not need
a change, and therefore, in some of the cases shown below, the ``n'' considered is higher (namely, 65 or 81), because it includes answers 
to those questions that were not modified after the first tests.

\vspace*{-0.5 cm}

\section{The instrument}

To elaborate the questions (Q) of our questionnaire, we used different sources which have studied in the past the most frequent answers of
preservice elementary teachers on topics of astronomy. A short list of references is (Callison \& Wright, 1993; Camino, 1995; Atwood \&
Atwood, 1995; Parker \& Heywood, 1998; 
Trumper, 2003; Mart{\'\i}nez-Sebasti\`a, 2004; Frede, 2006). 
Here we briefly describe our questions (more details are provided below):  
Q1 and Q2 ask about verticality and gravitation;  
Q3, Q4 and Q5 are about simple astronomical concepts. The fifth question pretends to analyze the consistency of the third answer;
Q6, Q7 and Q8 ask about the movement of the Moon and the three-body system (Sun, Earth and Moon); 
Q9  concerns the real movements of the Earth. It is directly connected with the third, fourth and fifth previous questions; 
Q10 asks about the Moon phases and it is connected with Q8; finally, 
Q11 and Q12 allow us to tests about cultural aspects of basic astronomy.  

\vspace*{-0.5 cm}

\section{First analysis of the results}

We present a summary of the first results, derived from the analysis of the final version of the instrument. We show them in percent
tables for a better interpretation. However, it must be said that these, for the time being, do not have statistical value. In the lists
below we show in shaded lines what we consider the expected (scientific) answers (for Q2 we allowed two answers to be the {\it correct}
ones, provided the justification offered by the students were adequate).

\smallskip
\noindent{\underline{\it Questions that ask about: verticality (Q1) and gravitation (Q2)}}

\begin{figure}[h!tb]
%\begin{figure}[b]
%\vspace*{-2.0 cm}
\begin{center} \includegraphics[width=10cm]{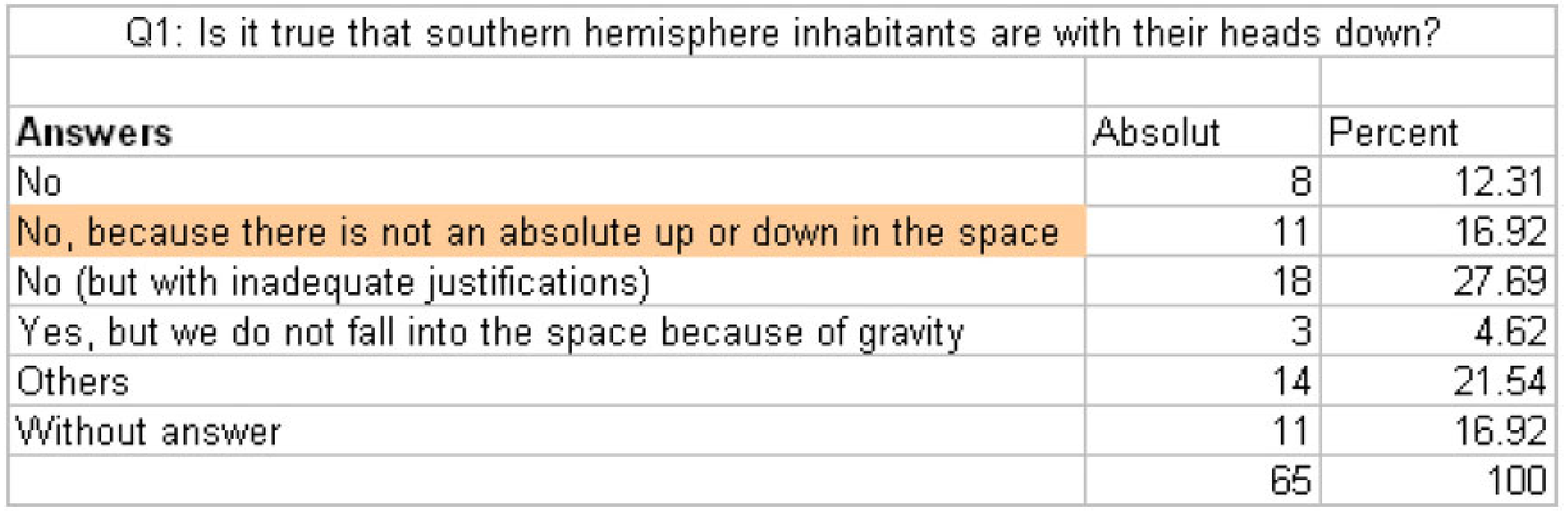} 
% \vspace*{-1.0 cm}
 \caption{Answers corresponding to prospective teachers' ideas on verticality.} \label{fig1}
\end{center}\end{figure}

About Q1, we note that the kind of answers obtained with preservice elementary teachers reflected many notions already present among
children (Nussbaum, 1979). Likewise, a 17\% responded according to the accepted scientific model, while a 12\% just answered ``No''
without further explanations and a 28\% answered the same but with inadequate reasonings.

\begin{figure}[h!tb]
\begin{center} 
\includegraphics[width=7.2cm]{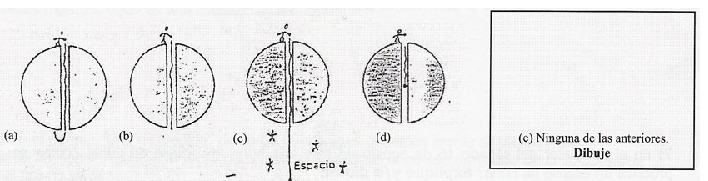}
\includegraphics[width=6.2cm]{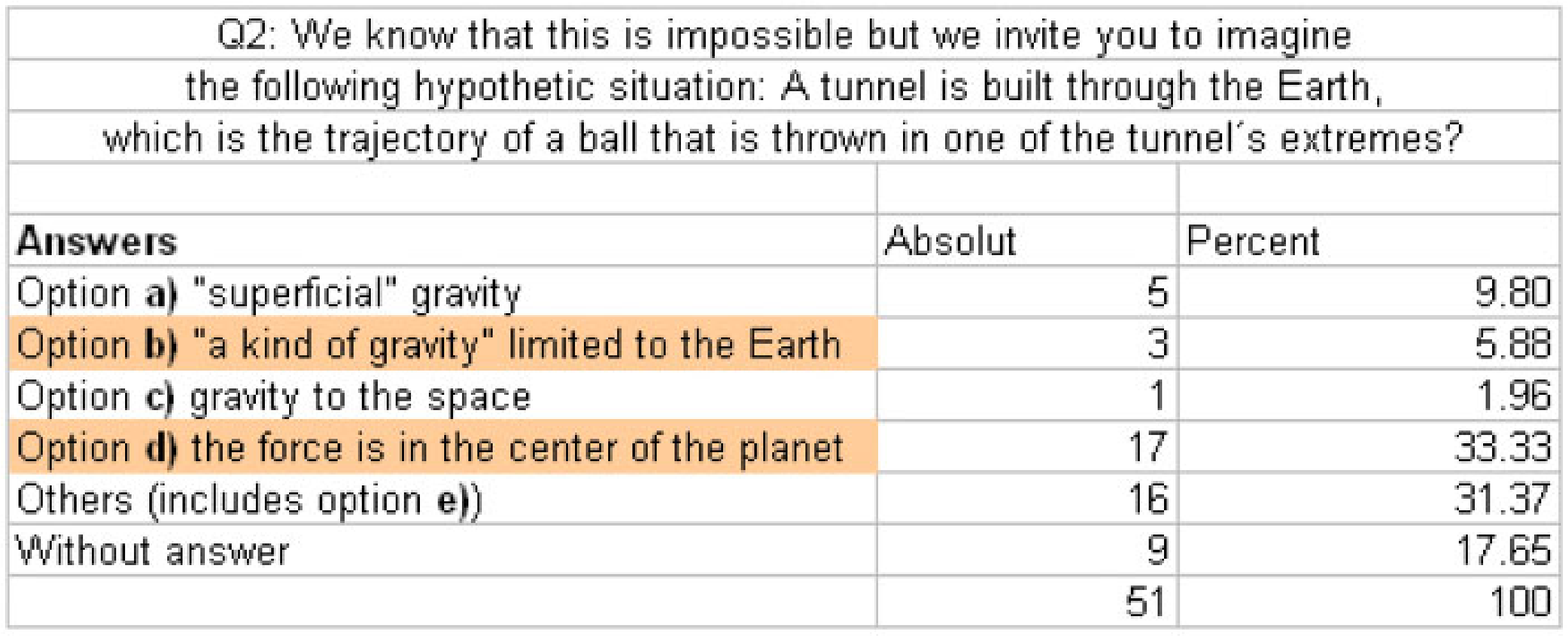}
\caption{Question and answers related to gravitation.} \label{fig2}\end{center}\end{figure}

Regarding Q2 one can say that many answers informed some notion of gravitation. They are in some stage between a primitive conception and
an actual scientific conception (cf. Nussbaum's analysis with children). We also point out that one of the inquired students presented a
notion very close to the scientific one (namely, the oscillation of the ball back and forth around the center of the Earth -- an option we
did not include in the test): choosing option (d) he then added ``[the ball] would stay at the center. Or may be it would reach the other
side but afterwards it will come back to the center''.

\smallskip
\noindent{\underline{\it Questions about simple astronomical concepts (Q3, Q4 and Q5)}}

\begin{figure}[h!tb] 
\begin{center}
\includegraphics[width=6.85cm]{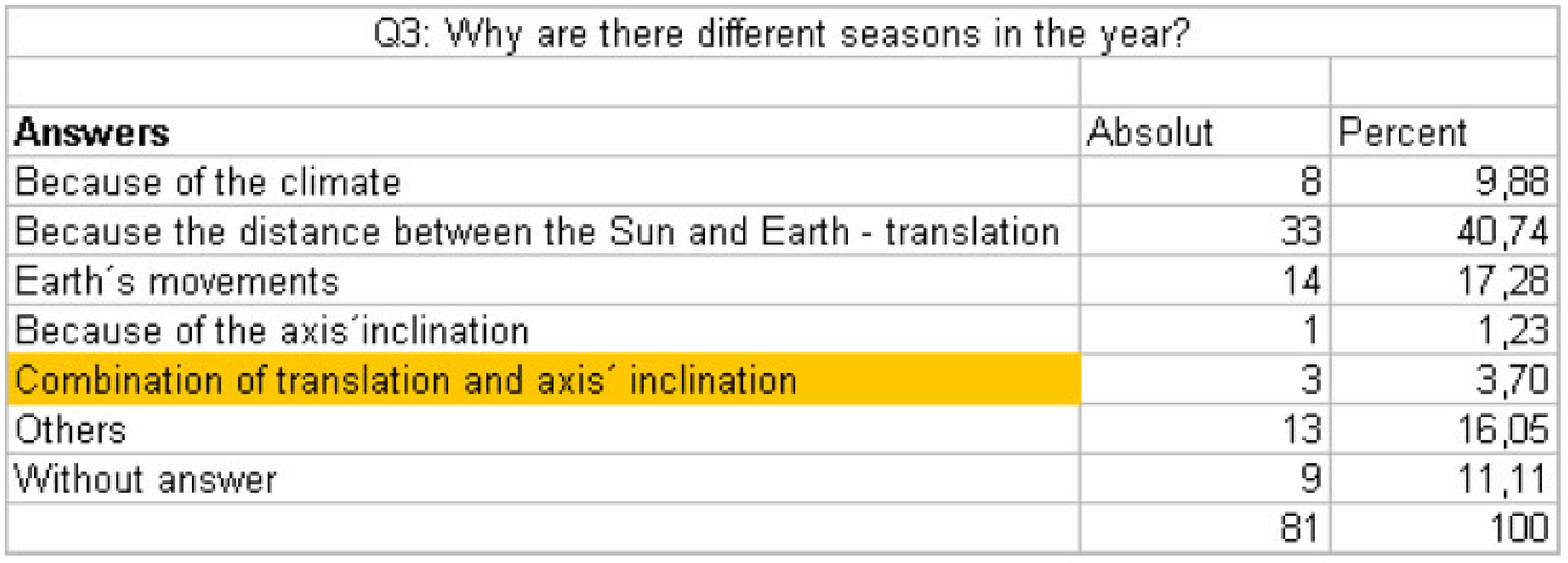}
\includegraphics[width=6.55cm]{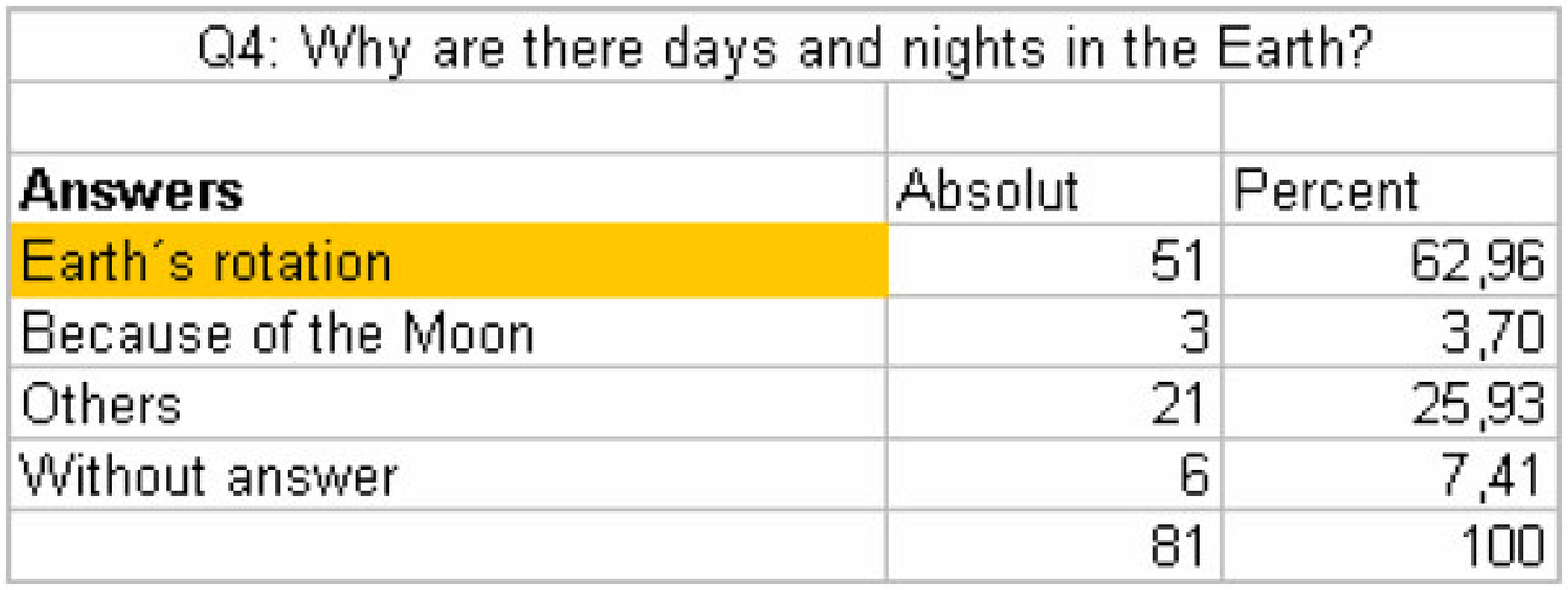}
\caption{Answers to some simple astronomical concepts.} \label{fig3-4}\end{center}\end{figure}

\begin{figure}[h!tb]
\begin{center} 
\includegraphics[width=6.3cm]{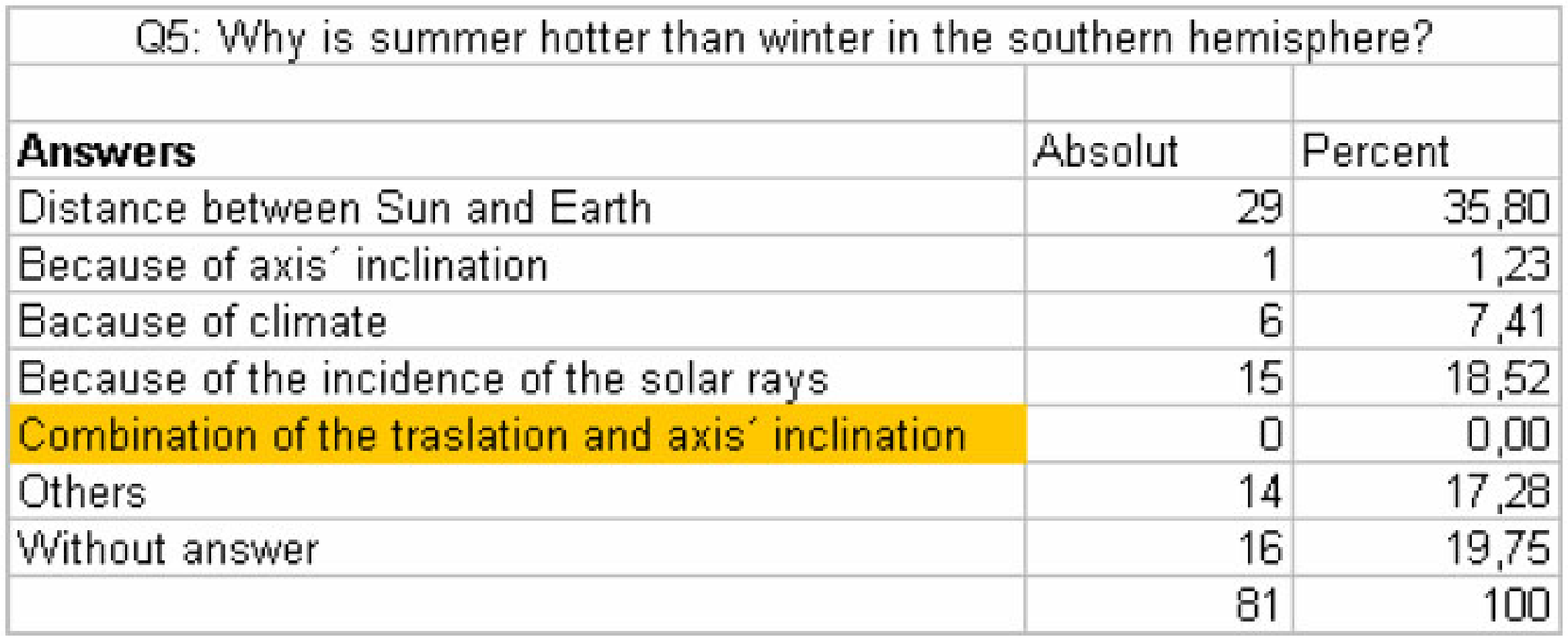}
\includegraphics[width=7.1cm]{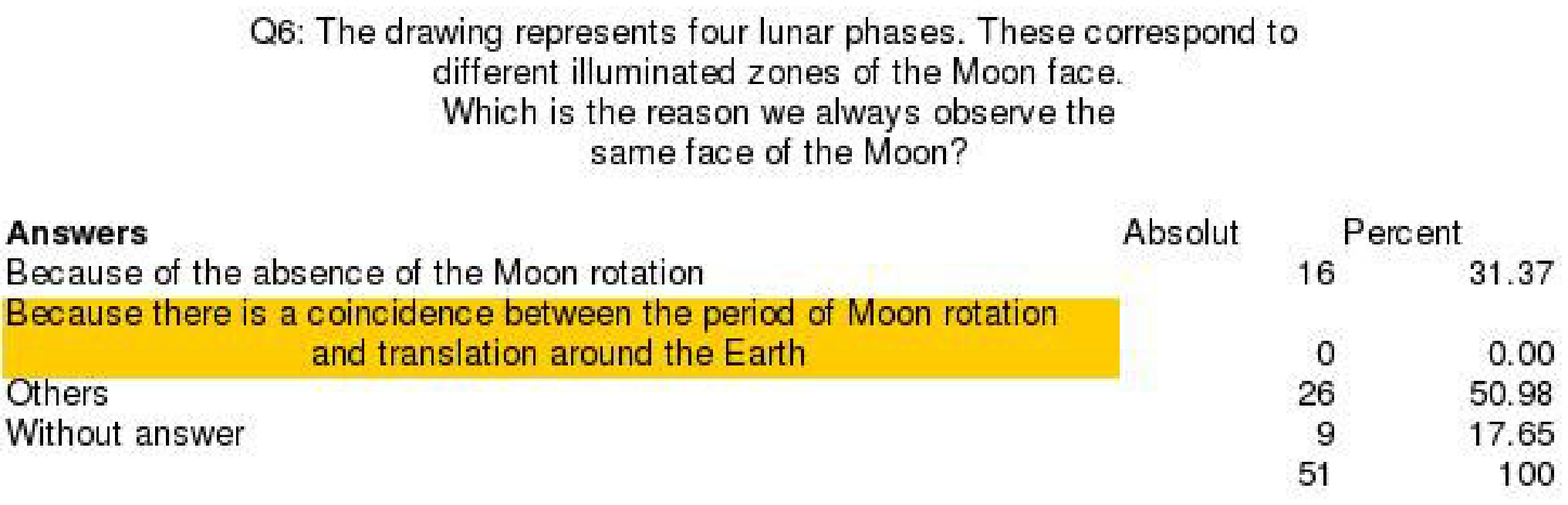}
\caption{Answers to questions about the seasons (left) and the face of the Moon (right).} 
\label{fig5-6}\end{center}\end{figure}

In Q3 (Fig.\,\ref{fig3-4}), nearly a 41\% of the total answers employed the so called {\it proximity theory} (which explains the existence
of seasons on the Earth using the wrong idea that our planet is closer to the Sun in summer than in winter). Linked to this, in Q5, a 36\%
made use of the same explanation (Fig.\,\ref{fig5-6}, left panel).  In this fifth question there was no complete adequate answer; however,
we can consider that ``because of the incidence of the solar rays'' goes in the right direction (18.5\% of the answers were like this).

Most of the answers to Q4 were correct (63\%). However, in most of them, translation and rotation of the planet were used interchangeably,
sometimes to describe the same phenomenon. Others, directly refer to ``circular movements'' or ``turn of the bodies'' to explain their
answers.

\smallskip
\noindent{\underline{\it Question on the Moon's movements and on the three-body system (Q6, Q7, Q8, Q10)}}

\begin{figure}[h!tb]
\begin{center} 
\includegraphics[width=6.9cm]{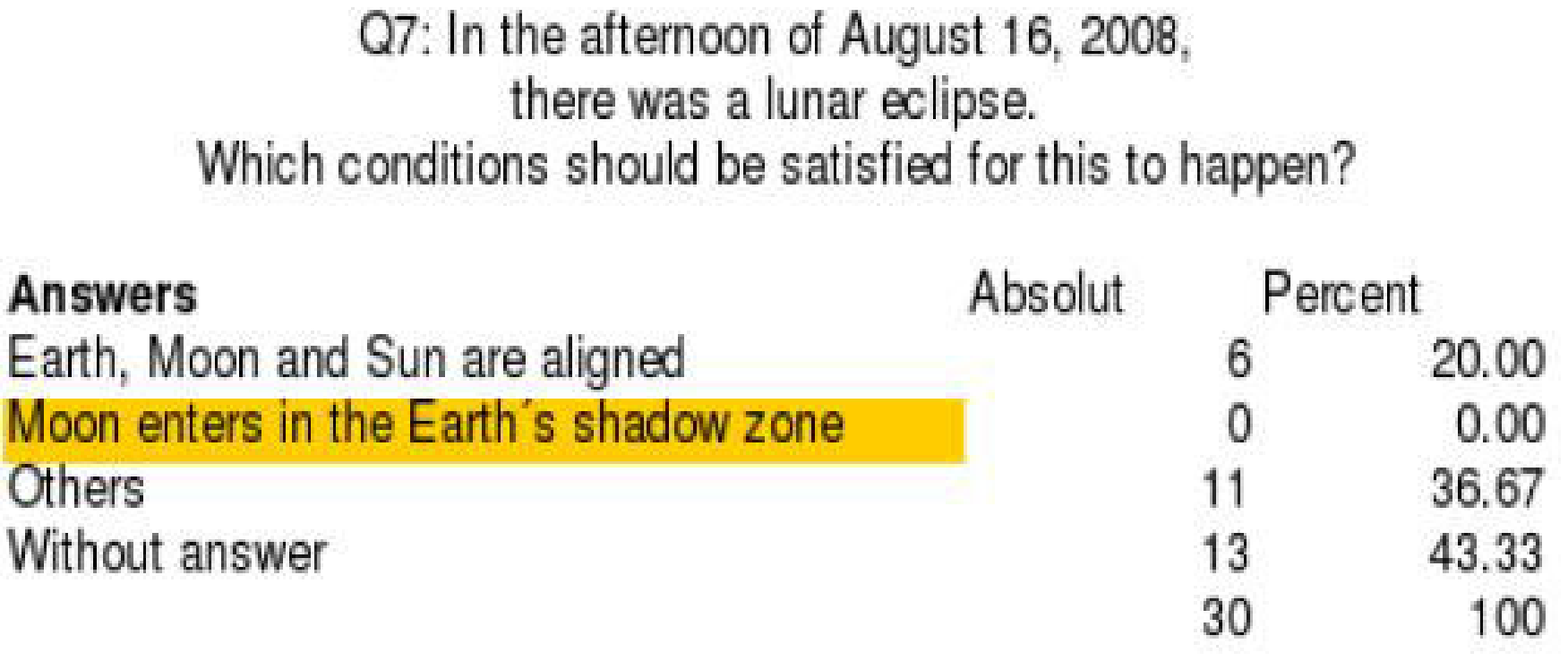}
\includegraphics[width=6.5cm]{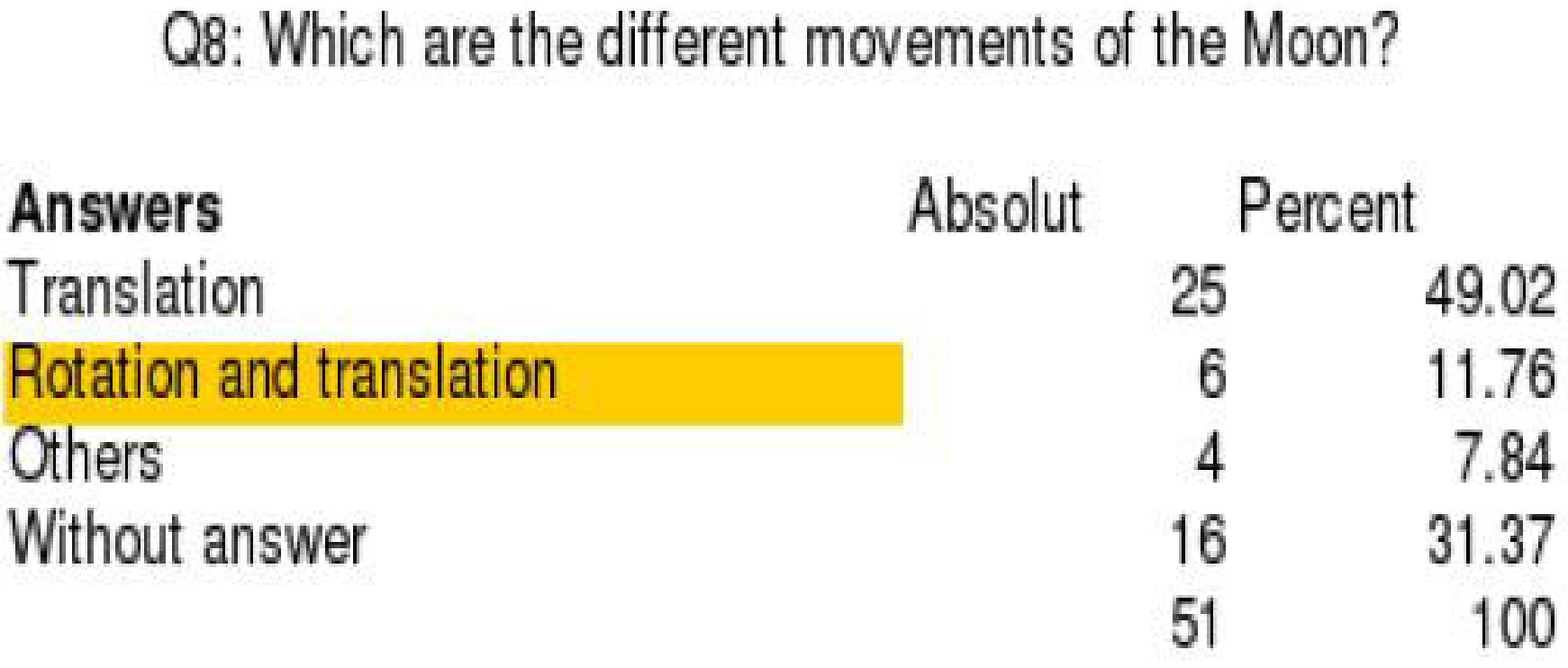}
\caption{Pilot test for the lunar eclipse (n=30 answers, left). Movements of the Moon (right).} \label{fig7-8}\end{center}\end{figure}

In Q8, most of the inquired students chose to answer that the Moon does not rotate on its axis (Fig.\,\ref{fig7-8}, right panel); for them
it just translates through the sky (nearly 50\% of the responses). This reasoning makes it difficult for them to explain other phenomena
related with the special motion of our satellite, as for example the observational fact that the Moon always points the same half of it
(its same {\it face}) towards the Earth. In fact, there was no correct answer to Q6 on this topic (see Fig.\,\ref{fig5-6}, right panel).

Moreover, none of the students could give a detailed correct explanation for the origin of a lunar eclipse (Q7), apart from just choosing
the option that there should be an alignment among the three celestial bodies (Fig.\,\ref{fig7-8}, left panel) which is also true in the
case of a Solar eclipse.

Question 10 was: {\it What is the origin of the Moon phases? Match one of the options. Explain your choice with a drawing}. We then
offered the following four choices: 
{\it (a)} The clouds cover a portion of the Moon so we only see the illuminated zone; 
{\it (b)} The Sun places itself between the Earth and the Moon, and projects a shadow; 
{\it (c)} The Moon reflects the light from the Sun, and we only see the illuminated part of the Moon; 
{\it (d)} The Sun illuminates the Moon but the Earth, in its movement, places itself between these bodies projecting a shadow on the Moon. 
Now, in the answers of this multiple choice Q10, more than 50\% of the people selected the correct option (c), although the drawings
provided as justifications were not always correct. Many of the wrong answers, as we expected, invoked some variation of the 
{\it eclipse theory} (option d). 

\smallskip
\noindent{\underline{\it Earth's movements (Q9)}}

In the first pilot tests, there was one question related to the movements of the Earth, expressed this way: 
{\it Explain what is the meaning of the rotation and translation movements of the Earth?} The answers we obtained, however, were somewhat
disappointing, as just 10\% of the inquired students got it right. For the final test we decided to split this question into two, asking
separately {\it What do we call the rotation of the Earth?} and {\it What do we call the translation of the Earth?} This simple
modification of the question changed completely the amount of correct answers. Out of 51 people we got roughly 40 whose answers agreed
with the present scientific model (approximately 80\%). 

\smallskip
\noindent{\underline{\it Definitions and astronomy in the daily language (Q11 and Q12)}}

Finally, we included a couple of questions related with astronomical elements that make part of our ordinary language, like the origin of
a shooting star and the real identity of the morning (or evening) star, planet Venus, which in many Spanish speaking countries is called
the {\it lucero}. So, our last two questions were Q11: {\it What is a shooting star?} and Q12: 
{\it In Jorge Luis Borges' poem ``A farewell'' (``Una despedida'') we can read the following: ``La noche hab{\'\i}a llegado con 
urgencia. Fuimos hasta la verja en esa gravedad de la sombra que ya el lucero alivia.'' What does Borges mean by ``the lucero''?}

For these two questions we obtained a long menu of answers, most of them wrong, e.g.: ``dead stars falling down in the atmosphere'' (Q11)
or ``shining objects'' either astronomical or otherwise (Q12). Just 10 out of 51 answers (roughly 20\%) got right Q11 on the explanation
of a shooting star. However, only 3 out of 51 pointed at planet Venus as the {\it lucero} which was mentioned by the famous Argentine
poet.

\vspace*{-0.5 cm}

\section{Final remarks}

We have tested some misconceptions present in preservice elementary teachers and compared them with the actual scientific notions. It is
surprising (or may be not) that quite a large proportion of these future teachers do not have the necessary knowledge that would entitle
them to cope with basic astronomical subjects in their own classes. We remark that although elementary school programmes include many of
these topics, appropriate training in astronomical subjects is in general lacking from teacher's schools. An appropriate design of
didactic tools that take into account teachers' misconceptions and other learning obstacles is currently under way.

\vspace*{-0.5 cm}

% see reviews by \cite[Anders \& Zinner (1993)]{AndersZinner93} and \cite[Ott (1993)]{Ott93}. 

\end{document}